\documentclass[aps,prl,superscriptaddress,10pt,notitlepage]{revtex4-1}

\usepackage{graphicx}%
\usepackage{dcolumn}%
\usepackage{bm}%

\begin{document}

\title{Doping Mechanisms in Graphene-MoS$_2$ Hybrids}

\author{B. Sachs}

\affiliation{I. Institut f{\"u}r Theoretische Physik, Universit{\"a}t Hamburg, Jungiusstra{\ss}e 9, D-20355 Hamburg, Germany}

\email{bsachs@physnet.uni-hamburg.de}

\author{L. Britnell}

\affiliation{School of Physics and Astronomy, University of Manchester, M13 9PL Manchester, United Kingdom}

\author{T. O. Wehling}

\affiliation{Institut f{\"u}r Theoretische Physik, Universit{\"a}t Bremen, Otto-Hahn-Allee 1, D-28359 Bremen, Germany}

\affiliation{Bremen Center for Computational Materials Science, Universit{\"a}t Bremen, Am Fallturm 1a, D-28359 Bremen, Germany}

\author{A. Eckmann}

\affiliation{School of Physics and Astronomy, University of Manchester, M13 9PL Manchester, United Kingdom}

\author{R. Jalil}

\affiliation{Manchester Centre for Mesoscience and Nanotechnology, University of Manchester, Manchester M13 9PL, United Kingdom}

\author{B. D. Belle}

\affiliation{Manchester Centre for Mesoscience and Nanotechnology, University of Manchester, Manchester M13 9PL, United Kingdom}

\author{A. I. Lichtenstein}

\affiliation{I. Institut f{\"u}r Theoretische Physik, Universit{\"a}t Hamburg, Jungiusstra{\ss}e 9, D-20355 Hamburg, Germany}

\author{M. I. Katsnelson}

\affiliation{Radboud University of Nijmegen, Institute for Molecules and Materials, Heijendaalseweg 135, 6525 AJ Nijmegen, The Netherlands}

\author{K. S. Novoselov}

\affiliation{School of Physics and Astronomy, University of Manchester, M13 9PL, Manchester, United Kingdom}

\begin{abstract}
We present a joint theoretical and experimental investigation of charge doping and electronic potential landscapes in hybrid structures composed of graphene and semiconducting single layer MoS$_2$. From first-principles simulations we find electron doping of graphene due to the presence of rhenium impurities in MoS$_2$. Furthermore, we show that MoS$_2$ edges give rise to charge reordering and a potential shift in graphene, which can be controlled through external gate voltages. The interplay of edge and impurity effects allows the use of the graphene-MoS$_2$ hybrid as a photodetector. Spatially resolved photocurrent signals can be used to resolve potential gradients and local doping levels in the sample.
\end{abstract}

\maketitle

Being a truly two-dimensional material \cite{art:2dimcrystals}, graphene can be integrated into hybrid structures with other 2D crystals such as boron nitride (BN), tungsten disulfide (WS$_2$) or molybdenum disulfide (MoS$_2$) \cite{ponomarenko2011tunable, britnell2012field,highkappa,dean2010boron,haigh2012cross}. The ability to build 'on demand' complex heterostructures via layer-by-layer integration establishes a whole family of new materials with widely varying characteristics and exciting possibilities for novel 2D nanodevices \cite{NovoselovNobel,grigorieva2013van,art:TMCD_review}. A prerequisite for future electronic applications lies in the understanding of interface effects when different building blocks come together. In particular, the electronic properties of realistic interfaces of graphene and two-dimensional materials present an open problem.  

In this work, we investigate heterostructures made of graphene and the semiconducting transition metal dichalcogenide MoS$_2$, a system that has already been utilized for vertical field-effect transistors \cite{britnell2012field}. We study how different charge transfer mechanisms control relative Fermi level positions, built-in electric fields and charge reordering at realistic graphene-MoS2 interfaces. The simulated charge and potential landscapes are compared to photovoltaic measurements. 

\begin{figure}
\includegraphics[width=0.69\linewidth]{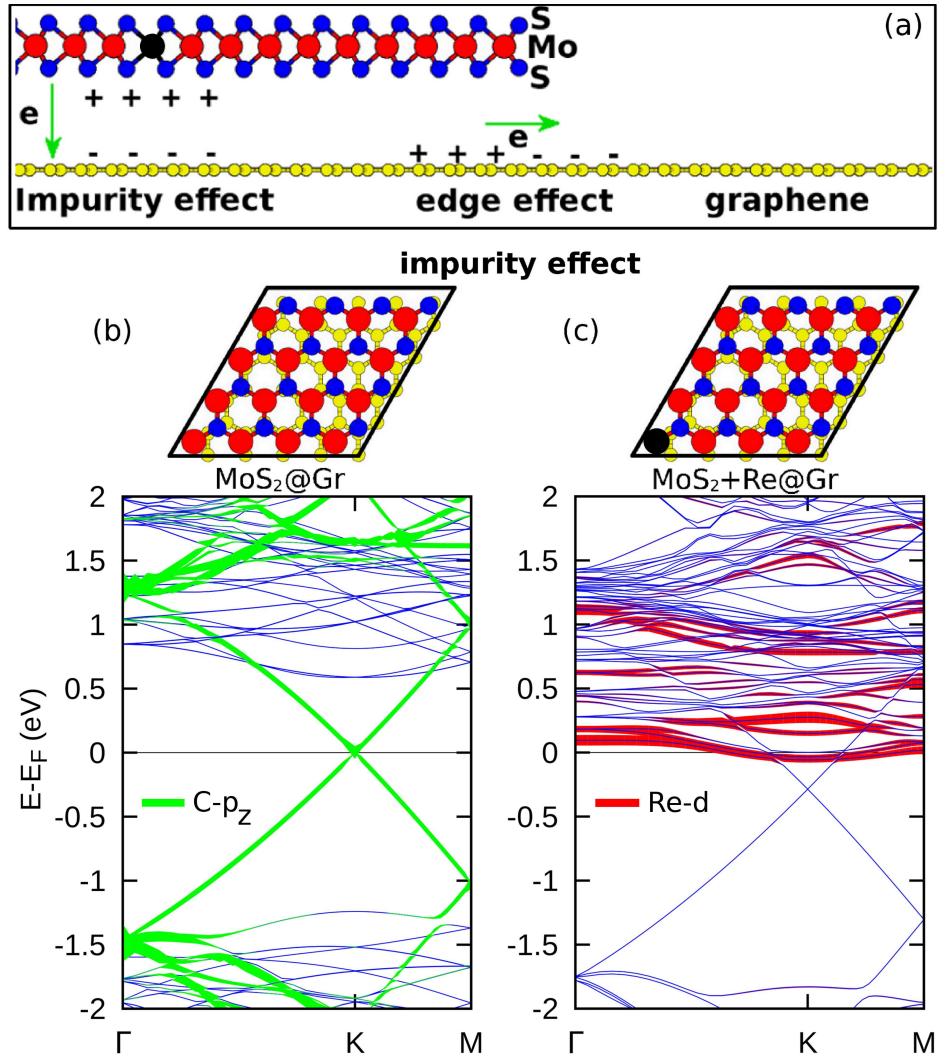}
\caption{(a) Sketch of the graphene-MoS$_2$ system from side view with visualization of planar and impurity doping mechanisms. (b) (upper panel) Unit cell of graphene fully covered with MoS$_2$ from top view.  Yellow atoms indicate the graphene carbon atoms, blue atoms sulfur and red atoms molybdenum. The stacking is chosen to be such that one C atom in the unit cell sits exactly below a Mo atom. In Ref. \cite{nanoscaleMoS2} it is stated that another configuration, where C sits below an S atom, is equivalent in both the binding and electronic properties. (lower panel) Band structure of the system. The green thick bands visualize the graphene p$_{\rm z}$ bands. (c) The same system with Re impurities (black atom in the unit cell). The red thick bands visualize contributions of the Re d orbitals to the band structure.}
\label{fig:bands}
\end{figure}

In order to investigate the graphene-MoS$_2$ hybrid structures theoretically, we performed first-principles density functional theory (DFT) simulations 
\footnote[1]{See supplementary material below for computational details, an extended discussion of different impurities and edge configurations as well as details about sample preparation.}. As the lattice constant of isolated graphene is about 23\% smaller than the one of isolated MoS$_2$, we constructed a supercell consisting of a 5x5 layer graphene (50 C atoms) coated with a 4x4 layer MoS$_2$ (16 Mo atoms and 32 S atoms) with a stacking as shown in Fig. \ref{fig:bands}b to account for this lattice mismatch. The remaining lattice mismatch of about 3-4\% is reasonably small and compensated by a slight strain of graphene to the MoS$_2$ lattice constant. 

The graphene-MoS$_2$ structure was then fully relaxed, which leads to an equilibrium graphene-MoS$_2$ distance of 3.35 \AA~ in good agreement with Ref. \cite{nanoscaleMoS2} and indicates a weak interlayer bonding of van der Waals-type. Based on this setup, we simulated realistic edge and impurity effects on the electronic properties of graphene-MoS$_2$ hybrids. 

We first address the role of impurity effects. To this end, we consider the fully MoS$_2$-covered graphene as in Fig. \ref{fig:bands}b without impurities and compare it to the case with impurities in the MoS$_2$. The band diagram of the pristine system is also shown in Fig. \ref{fig:bands}b. The green bands mark the graphene $p_{\rm z}$ contributions, and we see that the characteristic Dirac cone of graphene is preserved - with the Fermi level lying directly on the Dirac point. So, the interaction between graphene and clean MoS$_2$ is weak and does not induce an ``intrinsic'' doping in the graphene \footnote[2]{The only effect of MoS$_2$ on the low-energy states we see in a close-up is a small band gap of less than 1 meV respectively 2 meV with Re impurities. This is supposed to be analogous to the case of graphene on boron nitride, where small finite band gaps can occur due to local mass terms in the moir\'{e} cell \cite{sachs2011hBN}. However, this gap is too small to limit electron mobility significantly.} unlike e.g., SiC substrates \cite{SiCdoping, kim2008origin}.

The situation can change when impurities are present in the MoS$_2$. We performed X-ray Photoelectron Spectroscopy (XPS) measurements in order to verify the presence of impurities in the heterostructure. We found a significant amount of oxygen which we show to be not important in \footnotemark[1]. However, it is known from the literature that Re impurities naturally occur in MoS$_2$ \cite{stellman1998encyclopaedia, Greenwooed1997chemistry}. To simulate the effect of Re impurities, one Mo atom in the unit cell was replaced by a Re atom (Fig. \ref{fig:bands}c). We found that Re atoms substituting Mo atoms virtually do not alter the atomic structure of the system. In particular, the graphene-MoS$_2$ binding length and binding energy remain nearly the same. However, we see interesting electronic effects in the band diagram. Here, the Fermi level gets pinned at the impurity level in MoS$_2$, which leads to electron doping of graphene. The case depicted here, corresponds to a high impurity concentration of $7.6$x$10^{13}$cm$^{-2}$ and shows a Fermi-level shift of 0.29~eV, which corresponds to an electron doping of $0.8$x$10^{13}$cm$^{-2}$ . For an impurity concentration of $1.9$x$10^{13}$cm$^{-2}$ (using a larger supercell) we find a shift of 0.27~eV. Thus, indicating that the doping saturates for impurity concentrations on the order of $10^{13}$cm$^{-2}$ which is a result of the linear density of states in graphene and the relative position of the impurity level in MoS$_2$. The shift is also robust under the influence of additional MoS$_2$ layers or temperature \footnotemark[1]. The Re $d$ states are responsible for the electron transfer as can be seen from the red-colored thick bands in Fig. \ref{fig:bands}c. These form broad donor levels and pin the Fermi level whereby the Dirac point becomes shifted toward the MoS$_2$ conduction bands. The hybridization between MoS$_2$ and graphene remains weak and keeps the graphene dispersion unaffected near the Dirac point. Thus, Re impurities within the MoS$_2$ provide the electrons for an n-type doping of graphene while we find that commonly present oxygen impurities or sulfur vacancies do not affect graphene \footnotemark[1].

\begin{figure}
\includegraphics[width=0.69\linewidth]{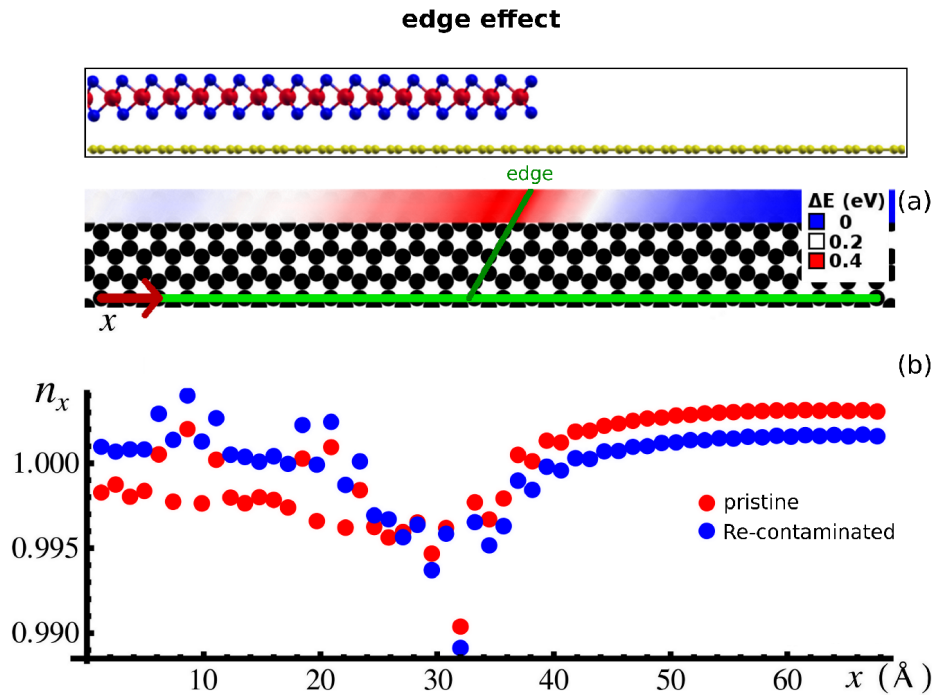}
\caption{(Black box) Side view on the MoS$_2$ (\=1010) edge configuration on graphene. (a) Visualization of the on-site energies in graphene under a MoS$_2$ edge for the case of pristine MoS$_2$ (black dots denote graphene atoms, Mo and S atoms are not shown for clarity). $\Delta E$ is defined as the energy relative to the minimum $p_{\rm z}$ energy (far away from the edge in uncovered graphene). (b) Occupation $n_x$ of graphene $p_{\rm z}$ orbitals along the green line in (a) for pristine (red) and Re-contaminated (blue) MoS$_2$ ($n_x>1$ means n-doping, $n_x<1$ p-doping; the max.  of $n_x=1.004$ corresponds to $1.5$x$10^{13}$cm$^{-2}$ carrier concentration).}  
\label{fig:step}
\end{figure}

For a more complete picture of the interface physics, we also consider charge redistributions within graphene in lateral direction below MoS$_2$ edges. Therefore, we have simulated graphene where one half of the graphene-plane is covered with MoS$_2$. A supercell containing about 600 atoms was constructed \footnotemark[1] whereby three realistic MoS$_2$ edge configurations \cite{edges1,edges2} were considered: the S-terminated (\=1010) edge and two Mo-terminated (10\=10) edges with additionally adsorbed S dimers or S monomers. Fig. \ref{fig:step} (black box) sketches the case of an S-terminated (\=1010) edge. 

\begin{figure}
\includegraphics[width=0.69\linewidth]{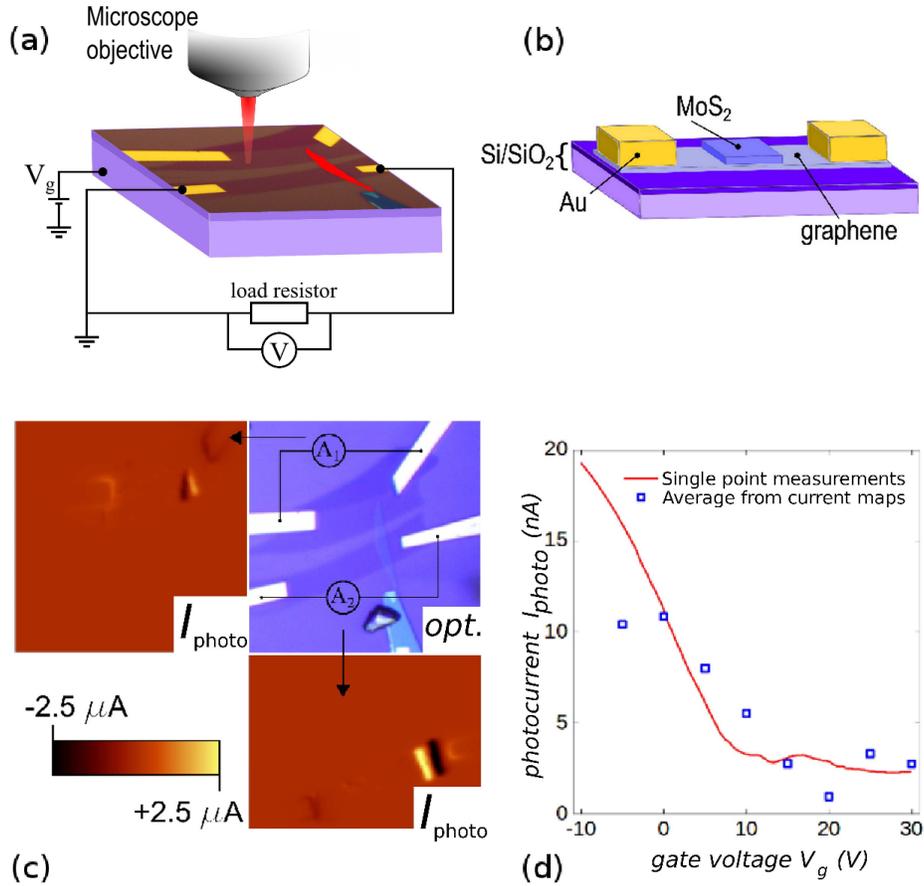}
\caption{(a) Sketch of the experimental setup for photocurrent measurements. Graphene (brown stripes) is partially covered with MoS$_2$ flakes (red stripe) with gold contacts (yellow) and irradiated by a laser beam). (b) Sketch of a lateral cut  through the sample for clarity. (c) Photocurrent maps with the terminals connected to the top (left top panel) or bottom (right bottom panel) electrodes. The dark blue stripes indicate graphene, the narrow light blue stripe in perpendicular direction denotes the MoS$_2$ stripe. (d) Dependence of the photocurrent $I_{\rm photo}$ from the gate voltage $V_{\rm g}$ from static point measurements (solid red line) and current maps averages (blue squares). } 	
\label{fig:Exp1}
\end{figure}

The onsite-energy differences $\Delta E$ of $p_{\rm z}$ orbitals in graphene atoms near this edge with pristine MoS$_2$ are visualized in Fig. \ref{fig:step}a.  A potential gradient at the edge (dark green line) can be observed with a maximum height difference of almost 400 meV. Similar to the case of graphene with metal contacts, which generally show a work function difference at the interface between graphene and metal/graphene regions \cite{Khomyakov_gr_metals}, an electric field builds up which results in charge separation, i.e. a p-n junction. We emphasize here that the shape of the potential gradient varies with the edge structure. For instance, we find the potential step to be smaller at the Mo-terminated (10\=10) edges. If impurities like Re are present in the MoS$_2$, the potential gradients created by work function differences at the edges are superimposed by bulk doping induced differences in the Fermi levels. This can be seen from the charge redistribution at this edge shown in Fig. \ref{fig:step}b. 

For both, pristine and Re-contaminated MoS$_2$, a zone of lower occupation occurs near the potential step at 32 \AA. In the pristine case (red dots) the uncovered graphene exhibits lower on-site energies and therefore gets locally n-doped while the MoS$_2$-covered part is p-doped with the exception of some single C atoms that sit below the middle of a MoS$_2$ hexagon (cf. Fig. \ref{fig:bands}b). This charge reordering is restricted to regions close to the edge, and under perfect conditions graphene is undoped on both sides at large distance. However, in a realistic system, Re impurities in the MoS$_2$ are present. Then the situation at the  (\=1010) edge is different: for a Re contamination as discussed above, the n-type bulk doping effect of MoS$_2$ on graphene and the edge-induced local p-doping compete. As a result, also the MoS$_2$-covered side becomes electron-doped in the edge region (Fig. \ref{fig:step}b, blue dots). 

To gain experimental insight into potential gradients at edges, we performed photocurrent measurements (Fig. \ref{fig:Exp1}a and b). Therefore, the sample was placed on a piezoelectric stage below a 1.96 eV (633 nm) laser. The current flowing in a circuit comprised the photoactive graphene - MoS$_2$ heterostructure and was measured as a function of the piezoelectric stage position \footnotemark[1]. The laser power was set at 80 $\mu$W and the laser spot radius was 0.5 $\mu$m. As has been previously reported \cite{park2009photocurrent}, there is a photovoltaic current generated at the interface between metal contacts and graphene due to p - n junctions. Similarly,  we measure a clear photocurrent signal when the laser spot is in the region where the MoS$_2$ has been placed (red region, Fig. \ref{fig:Exp1}a), while virtually no measurable current is generated in the uncovered graphene (purple region) and only a small current near the metal contacts (yellow regions). The measurements are shown in Fig. \ref{fig:Exp1}c and d. 
In \ref{fig:Exp1}c, photocurrent maps are recorded when the terminals are connected to either the top or bottom two electrodes respectively (top left and bottom right panel). The scanning photocurrent data was measured by recording the voltage drop across a load resistor of known resistance placed in series with the device. The response was determined for each step in the x-y position of the laser by measuring the voltage across this resistor. In this fashion, a spatial map of the local photocurrent distribution could be produced. In the top right panel an optical image of the device is shown. The photocurrent generated at the graphene - Au interface can be seen while a much stronger signal is obtained on the graphene - MoS$_2$ heterostructure. Also in some regions of uncovered graphene a small photocurrent is present due to small variations in local doping, but the effect is too minor to be visible on this scale. 

The symmetry of the photocurrent maps shows that the current is generated at the interface between the uncovered and the MoS$_2$-covered region. Since the uncovered graphene is undoped, while it is n-doped when covered with MoS$_2$, a potential barrier builds up as discussed above. The resulting electric field prevents a recombination of laser-induced electron-hole pairs and finally results in a current. Hence, it is the interplay of impurity doping and potential gradients that permits to use the graphene-MoS$_2$ heterostructure as a photodetector. Importantly, the current is not generated near the metal contacts like in most graphene-based photodetectors \cite{park2009photocurrent,xia2009photocurrent,echtermeyer2011strong,lee2008contact}. The photocurrent generated in a certain region of the sample is proportional to the local potential gradients. Therefore, the photocurrent map depicts local potential gradients and reveals the spatial dependence of the Fermi level. This technique is thus a powerful tool to investigate the doping levels in graphene heterostructures with spatial resolution.

The photocurrent signal $I_{\rm photo}$ of the active region responds to a gate voltage $V_{\rm g}$ (Fig. \ref{fig:Exp1}d), and approaches a small constant value at $\sim 30$~V which corresponds to an electron concentration of  $2$x$10^{12}$cm$^{-2}$. The response of $I_{\rm photo}$ to $V_{\rm g}$ was measured in two ways: i. the laser was statically positioned on the heterostructure (solid red line) while sweeping gate voltage. ii. photocurrent maps were taken at a series of different gate voltage values and the mean current from the active region calculated (blue squares). For these measurements a potential was applied between the silicon backgate and graphene, while the graphene was kept grounded. For negative gate voltages the photocurrent signal increased, which corresponds to an increased potential step at the edges. For positive gate voltages, the photocurrent signal decreased until the heterostructure region was barely distinguishable. Since a positive sign of $V_{\rm g}$ corresponds to an electron doping, this confirms the n-type doping of graphene under MoS$_2$. This is also in agreement with Ref. \cite{zhang2013ultrahigh}, where an n-type doping of graphene on MoS$_2$ is reported. 

\begin{figure}
\includegraphics[width=0.69\linewidth]{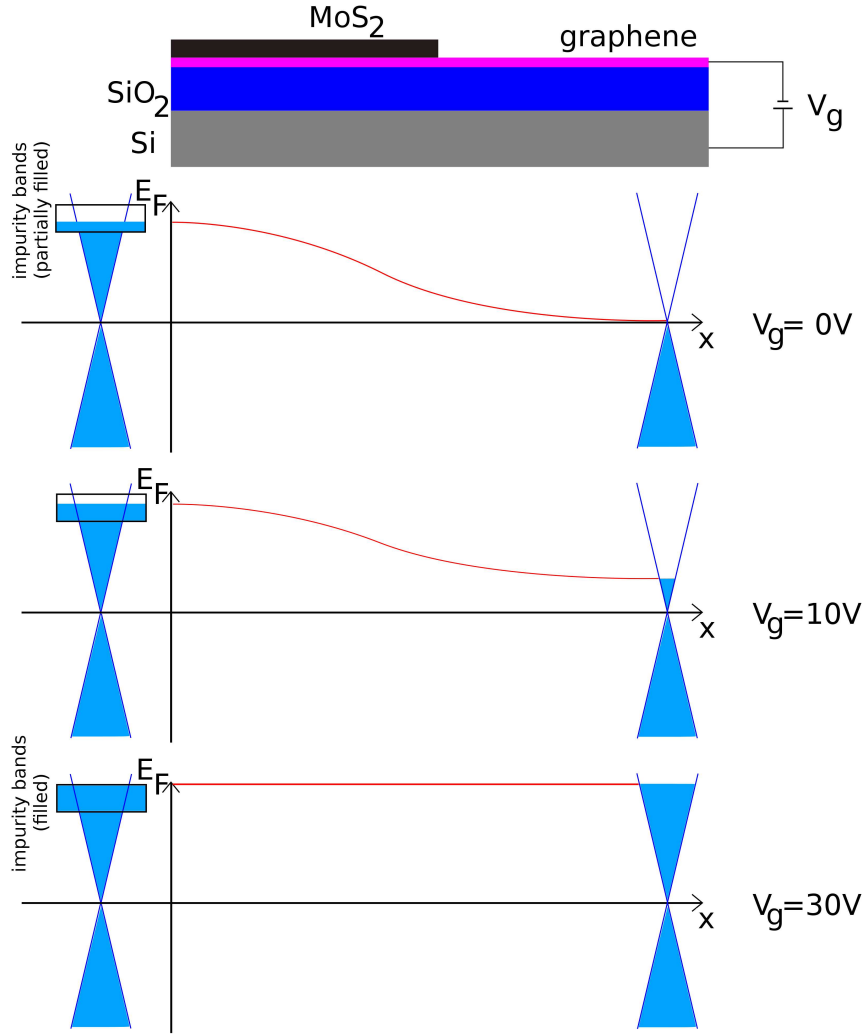}
\caption{Illustration of the Fermi level evolution along the MoS$_2$ edge for different gate voltages $V_{\rm g}$. In the MoS$_2$-covered region (left), the Fermi level becomes pinned due to impurity bands while it increases in the uncovered region (right) with $V_{\rm g}$.} 	
\label{fig:Dirac}
\end{figure}

Apparently, the potential barrier changes upon applying a gate voltage $V_{\rm g}$. The local Fermi level in MoS$_2$-covered and uncovered graphene responds differently  to $V_{\rm g}$. This is because the MoS$_2$ on graphene can partially screen the gate induced electric fields due to quantum capacitance related charge redistributions from graphene to MoS$_2$ (Fig. \ref{fig:Dirac}): in uncovered graphene, all charge density $\rho$ induced by $V_{\rm g}$ has to be taken up by the graphene bands. Thus, the Fermi level in graphene increases correspondingly to $\rho$. On the contrary, in MoS$_2$-covered graphene in addition to states derived from graphene, MoS$_2$ states or impurity states from MoS$_2$ can be available at the Fermi level. These latter MoS$_2$ and impurity-derived states take up large parts of the charge (due to the low density of states in graphene) and thus, the gate voltage-induced changes of the Fermi level are much smaller in the uncovered than in the covered region. In other words, impurity levels in MoS$_2$ induce a Fermi level pinning as discussed in Fig. \ref{fig:bands}c. This mechanism is responsible for high photocurrents at negative gate voltages: the uncovered graphene becomes hole-doped while the MoS$_2$-covered region remains with the Fermi level pinned to the impurity donor states. Thus, the potential step at the MoS$_2$ edges increases with negative gate voltages which amplifies the photocurrent. For positive gate voltages, the Fermi level in uncovered graphene increases and the potential step is reduced. At very large gate voltages, we speculate that the impurity donor bands become filled and there is essentially no charge transfer from the donor bands to graphene. Therefore, the potential step and the photocurrent generated at the MoS$_2$ edge become very small.

In conclusion, we presented an investigation of graphene-MoS$_2$ hybrid structures. Our DFT calculations (cf. Ref. \cite{sachs2011hBN} for graphene-BN hybrids) reveal two different doping mechanisms: First, we showed that MoS$_2$ edges induce charge redistributions within the graphene plane. Second, Re impurities in the MoS$_2$ lead to charge transfer from MoS$_2$ to the graphene. Photovoltaic experiments support these findings. The interplay of impurity as well as edge effects allows to build graphene-MoS$_2$ photodetector devices where the photocurrent is not generated at the metal contacts. 

Support from the Deut\-sche For\-schungs\-ge\-mein\-schaft via SPP 1459 and the European Graphene Flagship are acknowledged.


\begin{thebibliography}{25}%
\makeatletter
\providecommand \@ifxundefined [1]{%
 \@ifx{#1\undefined}
}%
\providecommand \@ifnum [1]{%
 \ifnum #1\expandafter \@firstoftwo
 \else \expandafter \@secondoftwo
 \fi
}%
\providecommand \@ifx [1]{%
 \ifx #1\expandafter \@firstoftwo
 \else \expandafter \@secondoftwo
 \fi
}%
\providecommand \natexlab [1]{#1}%
\providecommand \enquote  [1]{``#1''}%
\providecommand \bibnamefont  [1]{#1}%
\providecommand \bibfnamefont [1]{#1}%
\providecommand \citenamefont [1]{#1}%
\providecommand \href@noop [0]{\@secondoftwo}%
\providecommand \href [0]{\begingroup \@sanitize@url \@href}%
\providecommand \@href[1]{\@@startlink{#1}\@@href}%
\providecommand \@@href[1]{\endgroup#1\@@endlink}%
\providecommand \@sanitize@url [0]{\catcode `\\12\catcode `\$12\catcode
  `\&12\catcode `\#12\catcode `\^12\catcode `\_12\catcode `\%12\relax}%
\providecommand \@@startlink[1]{}%
\providecommand \@@endlink[0]{}%
\providecommand \url  [0]{\begingroup\@sanitize@url \@url }%
\providecommand \@url [1]{\endgroup\@href {#1}{\urlprefix }}%
\providecommand \urlprefix  [0]{URL }%
\providecommand \Eprint [0]{\href }%
\providecommand \doibase [0]{http://dx.doi.org/}%
\providecommand \selectlanguage [0]{\@gobble}%
\providecommand \bibinfo  [0]{\@secondoftwo}%
\providecommand \bibfield  [0]{\@secondoftwo}%
\providecommand \translation [1]{[#1]}%
\providecommand \BibitemOpen [0]{}%
\providecommand \bibitemStop [0]{}%
\providecommand \bibitemNoStop [0]{.\EOS\space}%
\providecommand \EOS [0]{\spacefactor3000\relax}%
\providecommand \BibitemShut  [1]{\csname bibitem#1\endcsname}%
\let\auto@bib@innerbib\@empty
%</preamble>
\bibitem [{\citenamefont {Novoselov}\ \emph {et~al.}(2005)\citenamefont
  {Novoselov}, \citenamefont {Jiang}, \citenamefont {Schedin}, \citenamefont
  {Booth}, \citenamefont {Khotkevich}, \citenamefont {Morozov},\ and\
  \citenamefont {Geim}}]{art:2dimcrystals}%
  \BibitemOpen
  \bibfield  {author} {\bibinfo {author} {\bibfnamefont {K.~S.}\ \bibnamefont
  {Novoselov}}, \bibinfo {author} {\bibfnamefont {D.}~\bibnamefont {Jiang}},
  \bibinfo {author} {\bibfnamefont {F.}~\bibnamefont {Schedin}}, \bibinfo
  {author} {\bibfnamefont {T.~J.}\ \bibnamefont {Booth}}, \bibinfo {author}
  {\bibfnamefont {V.~V.}\ \bibnamefont {Khotkevich}}, \bibinfo {author}
  {\bibfnamefont {S.~V.}\ \bibnamefont {Morozov}}, \ and\ \bibinfo {author}
  {\bibfnamefont {A.~K.}\ \bibnamefont {Geim}},\ } {\bibfield  {journal} {\bibinfo  {journal} {PNAS}\
  }\textbf {\bibinfo {volume} {102}},\ \bibinfo {pages} {10451} (\bibinfo
  {year} {2005})}\BibitemShut {NoStop}%
\bibitem [{\citenamefont {Ponomarenko}\ \emph {et~al.}(2011)\citenamefont
  {Ponomarenko}, \citenamefont {Geim}, \citenamefont {Zhukov}, \citenamefont
  {Jalil}, \citenamefont {Morozov}, \citenamefont {Grigorieva}, \citenamefont
  {Novoselov}, \citenamefont {Hill}, \citenamefont {Cheianov}, \citenamefont
  {Falko}, \citenamefont {Watanabe}, \citenamefont {Taniguchi},\ and\
  \citenamefont {Gorbachev}}]{ponomarenko2011tunable}%
  \BibitemOpen
  \bibfield  {author} {\bibinfo {author} {\bibfnamefont {L.~A.}\ \bibnamefont
  {Ponomarenko}}, \bibinfo {author} {\bibfnamefont {A.~K.}\ \bibnamefont
  {Geim}}, \bibinfo {author} {\bibfnamefont {A.~A.}\ \bibnamefont {Zhukov}},
  \bibinfo {author} {\bibfnamefont {R.}~\bibnamefont {Jalil}}, \bibinfo
  {author} {\bibfnamefont {S.~V.}\ \bibnamefont {Morozov}}, \bibinfo {author}
  {\bibfnamefont {I.~V.}\ \bibnamefont {Grigorieva}}, \bibinfo {author}
  {\bibfnamefont {K.~S.}\ \bibnamefont {Novoselov}}, \bibinfo {author}
  {\bibfnamefont {E.~H.}\ \bibnamefont {Hill}}, \bibinfo {author}
  {\bibfnamefont {V.~V.}\ \bibnamefont {Cheianov}}, \bibinfo {author}
  {\bibfnamefont {V.~I.}\ \bibnamefont {Falko}}, \bibinfo {author}
  {\bibfnamefont {K.}~\bibnamefont {Watanabe}}, \bibinfo {author}
  {\bibfnamefont {T.}~\bibnamefont {Taniguchi}}, \ and\ \bibinfo {author}
  {\bibfnamefont {R.~V.}\ \bibnamefont {Gorbachev}},\ }\href@noop {} {\bibfield
   {journal} {\bibinfo  {journal} {Nature Phys.}\ }\textbf {\bibinfo {volume}
  {7}},\ \bibinfo {pages} {958} (\bibinfo {year} {2011})}\BibitemShut {NoStop}%
\bibitem [{\citenamefont {Britnell}\ \emph {et~al.}(2012)\citenamefont
  {Britnell}, \citenamefont {Gorbachev}, \citenamefont {Jalil}, \citenamefont
  {Belle}, \citenamefont {Schedin}, \citenamefont {Mishchenko}, \citenamefont
  {Georgiou}, \citenamefont {Katsnelson}, \citenamefont {Eaves}, \citenamefont
  {Morozov}, \citenamefont {Peres}, \citenamefont {Leist}, \citenamefont
  {Geim}, \citenamefont {Novoselov},\ and\ \citenamefont
  {Ponomarenko}}]{britnell2012field}%
  \BibitemOpen
  \bibfield  {author} {\bibinfo {author} {\bibfnamefont {L.}~\bibnamefont
  {Britnell}}, \bibinfo {author} {\bibfnamefont {R.~V.}\ \bibnamefont
  {Gorbachev}}, \bibinfo {author} {\bibfnamefont {R.}~\bibnamefont {Jalil}},
  \bibinfo {author} {\bibfnamefont {B.~D.}\ \bibnamefont {Belle}}, \bibinfo
  {author} {\bibfnamefont {F.}~\bibnamefont {Schedin}}, \bibinfo {author}
  {\bibfnamefont {A.}~\bibnamefont {Mishchenko}}, \bibinfo {author}
  {\bibfnamefont {T.}~\bibnamefont {Georgiou}}, \bibinfo {author}
  {\bibfnamefont {M.~I.}\ \bibnamefont {Katsnelson}}, \bibinfo {author}
  {\bibfnamefont {L.}~\bibnamefont {Eaves}}, \bibinfo {author} {\bibfnamefont
  {S.~V.}\ \bibnamefont {Morozov}}, \bibinfo {author} {\bibfnamefont
  {N.~M.~R.}\ \bibnamefont {Peres}}, \bibinfo {author} {\bibfnamefont
  {J.}~\bibnamefont {Leist}}, \bibinfo {author} {\bibfnamefont {A.~K.}\
  \bibnamefont {Geim}}, \bibinfo {author} {\bibfnamefont {K.~S.}\ \bibnamefont
  {Novoselov}}, \ and\ \bibinfo {author} {\bibfnamefont {L.~A.}\ \bibnamefont
  {Ponomarenko}},\ }\href@noop {} {\bibfield  {journal} {\bibinfo  {journal}
  {Science}\ }\textbf {\bibinfo {volume} {335}},\ \bibinfo {pages} {947}
  (\bibinfo {year} {2012})}\BibitemShut {NoStop}%
\bibitem [{\citenamefont {Ponomarenko}\ \emph {et~al.}(2009)\citenamefont
  {Ponomarenko}, \citenamefont {Yang}, \citenamefont {Mohiuddin}, \citenamefont
  {Katsnelson}, \citenamefont {Novoselov}, \citenamefont {Morozov},
  \citenamefont {Zhukov}, \citenamefont {Schedin}, \citenamefont {Hill},\ and\
  \citenamefont {Geim}}]{highkappa}%
  \BibitemOpen
  \bibfield  {author} {\bibinfo {author} {\bibfnamefont {L.~A.}\ \bibnamefont
  {Ponomarenko}}, \bibinfo {author} {\bibfnamefont {R.}~\bibnamefont {Yang}},
  \bibinfo {author} {\bibfnamefont {T.~M.}\ \bibnamefont {Mohiuddin}}, \bibinfo
  {author} {\bibfnamefont {M.~I.}\ \bibnamefont {Katsnelson}}, \bibinfo
  {author} {\bibfnamefont {K.~S.}\ \bibnamefont {Novoselov}}, \bibinfo {author}
  {\bibfnamefont {S.~V.}\ \bibnamefont {Morozov}}, \bibinfo {author}
  {\bibfnamefont {A.~A.}\ \bibnamefont {Zhukov}}, \bibinfo {author}
  {\bibfnamefont {F.}~\bibnamefont {Schedin}}, \bibinfo {author} {\bibfnamefont
  {E.~W.}\ \bibnamefont {Hill}}, \ and\ \bibinfo {author} {\bibfnamefont
  {A.~K.}\ \bibnamefont {Geim}},\ }\href@noop {} {\bibfield  {journal}
  {\bibinfo  {journal} {Phys. Rev. Lett.}\ }\textbf {\bibinfo {volume} {102}},\
  \bibinfo {pages} {206603} (\bibinfo {year} {2009})}\BibitemShut {NoStop}%
\bibitem [{\citenamefont {Dean}\ \emph {et~al.}(2010)\citenamefont {Dean},
  \citenamefont {Young}, \citenamefont {Meric}, \citenamefont {Lee},
  \citenamefont {Wang}, \citenamefont {Sorgenfrei}, \citenamefont {Watanabe},
  \citenamefont {Taniguchi}, \citenamefont {Kim}, \citenamefont {Shepard},\
  and\ \citenamefont {Hone}}]{dean2010boron}%
  \BibitemOpen
  \bibfield  {author} {\bibinfo {author} {\bibfnamefont {C.~R.}\ \bibnamefont
  {Dean}}, \bibinfo {author} {\bibfnamefont {A.~F.}\ \bibnamefont {Young}},
  \bibinfo {author} {\bibfnamefont {I.}~\bibnamefont {Meric}}, \bibinfo
  {author} {\bibfnamefont {C.}~\bibnamefont {Lee}}, \bibinfo {author}
  {\bibfnamefont {L.}~\bibnamefont {Wang}}, \bibinfo {author} {\bibfnamefont
  {S.}~\bibnamefont {Sorgenfrei}}, \bibinfo {author} {\bibfnamefont
  {K.}~\bibnamefont {Watanabe}}, \bibinfo {author} {\bibfnamefont
  {T.}~\bibnamefont {Taniguchi}}, \bibinfo {author} {\bibfnamefont
  {P.}~\bibnamefont {Kim}}, \bibinfo {author} {\bibfnamefont {K.~L.}\
  \bibnamefont {Shepard}}, \ and\ \bibinfo {author} {\bibfnamefont
  {J.}~\bibnamefont {Hone}},\ }\href@noop {} {\bibfield  {journal} {\bibinfo
  {journal} {Nature Nanotech.}\ }\textbf {\bibinfo {volume} {5}},\ \bibinfo
  {pages} {722} (\bibinfo {year} {2010})}\BibitemShut {NoStop}%
\bibitem [{\citenamefont {Haigh}\ \emph {et~al.}(2012)\citenamefont {Haigh},
  \citenamefont {Gholinia}, \citenamefont {Jalil}, \citenamefont {Romani},
  \citenamefont {Britnell}, \citenamefont {Elias}, \citenamefont {Novoselov},
  \citenamefont {Ponomarenko}, \citenamefont {Geim},\ and\ \citenamefont
  {Gorbachev}}]{haigh2012cross}%
  \BibitemOpen
  \bibfield  {author} {\bibinfo {author} {\bibfnamefont {S.~J.}\ \bibnamefont
  {Haigh}}, \bibinfo {author} {\bibfnamefont {A.}~\bibnamefont {Gholinia}},
  \bibinfo {author} {\bibfnamefont {R.}~\bibnamefont {Jalil}}, \bibinfo
  {author} {\bibfnamefont {S.}~\bibnamefont {Romani}}, \bibinfo {author}
  {\bibfnamefont {L.}~\bibnamefont {Britnell}}, \bibinfo {author}
  {\bibfnamefont {D.~C.}\ \bibnamefont {Elias}}, \bibinfo {author}
  {\bibfnamefont {K.~S.}\ \bibnamefont {Novoselov}}, \bibinfo {author}
  {\bibfnamefont {L.~A.}\ \bibnamefont {Ponomarenko}}, \bibinfo {author}
  {\bibfnamefont {A.~K.}\ \bibnamefont {Geim}}, \ and\ \bibinfo {author}
  {\bibfnamefont {R.}~\bibnamefont {Gorbachev}},\ }\href@noop {} {\bibfield
  {journal} {\bibinfo  {journal} {Nature Materials}\ }\textbf {\bibinfo
  {volume} {11}},\ \bibinfo {pages} {764} (\bibinfo {year} {2012})}\BibitemShut
  {NoStop}%
\bibitem [{\citenamefont {Novoselov}(2011)}]{NovoselovNobel}%
  \BibitemOpen
  \bibfield  {author} {\bibinfo {author} {\bibfnamefont {K.~S.}\ \bibnamefont
  {Novoselov}},\ } {\bibfield
  {journal} {\bibinfo  {journal} {Rev. Mod. Phys.}\ }\textbf {\bibinfo {volume}
  {83}},\ \bibinfo {pages} {837} (\bibinfo {year} {2011})}\BibitemShut
  {NoStop}%
\bibitem [{\citenamefont {Geim}\ and\ \citenamefont
  {Grigorieva}(2013)}]{grigorieva2013van}%
  \BibitemOpen
  \bibfield  {author} {\bibinfo {author} {\bibfnamefont {A.~K.}\ \bibnamefont
  {Geim}}\ and\ \bibinfo {author} {\bibfnamefont {I.~V.}\ \bibnamefont
  {Grigorieva}},\ }\href@noop {} {\bibfield  {journal} {\bibinfo  {journal} {Nature}\
  }\textbf {\bibinfo {volume} {499}},\ \bibinfo {pages} {419} (\bibinfo {year}
  {2013})}\BibitemShut {NoStop}%
\bibitem [{\citenamefont {Wang}\ \emph {et~al.}(2012)\citenamefont {Wang},
  \citenamefont {Kalantar-Zadeh}, \citenamefont {Kis}, \citenamefont
  {Coleman},\ and\ \citenamefont {Strano}}]{art:TMCD_review}%
  \BibitemOpen
  \bibfield  {author} {\bibinfo {author} {\bibfnamefont {Q.~H.}\ \bibnamefont
  {Wang}}, \bibinfo {author} {\bibfnamefont {K.}~\bibnamefont
  {Kalantar-Zadeh}}, \bibinfo {author} {\bibfnamefont {A.}~\bibnamefont {Kis}},
  \bibinfo {author} {\bibfnamefont {J.~N.}\ \bibnamefont {Coleman}}, \ and\
  \bibinfo {author} {\bibfnamefont {M.~S.}\ \bibnamefont {Strano}},\
  }\href@noop {} {\bibfield  {journal} {\bibinfo  {journal} {Nature Nanotech.}\
  }\textbf {\bibinfo {volume} {7}},\ \bibinfo {pages} {699} (\bibinfo {year}
  {2012})}\BibitemShut {NoStop}%
\bibitem [{Note1()}]{Note1}%
  \BibitemOpen
  \bibinfo {note} {See supplementary material at [url] for computational
  details, an extended discussion of different impurities and edge
  configurations as well as details about sample preparation.}\BibitemShut
  {Stop}%
\bibitem [{\citenamefont {Ma}\ \emph {et~al.}(2011)\citenamefont {Ma},
  \citenamefont {Dai}, \citenamefont {Guo}, \citenamefont {Niu},\ and\
  \citenamefont {Huang}}]{nanoscaleMoS2}%
  \BibitemOpen
  \bibfield  {author} {\bibinfo {author} {\bibfnamefont {Y.}~\bibnamefont
  {Ma}}, \bibinfo {author} {\bibfnamefont {Y.}~\bibnamefont {Dai}}, \bibinfo
  {author} {\bibfnamefont {M.}~\bibnamefont {Guo}}, \bibinfo {author}
  {\bibfnamefont {C.}~\bibnamefont {Niu}}, \ and\ \bibinfo {author}
  {\bibfnamefont {B.}~\bibnamefont {Huang}},\ }\href@noop {} {\bibfield
  {journal} {\bibinfo  {journal} {Nanoscale}\ }\textbf {\bibinfo {volume}
  {3}},\ \bibinfo {pages} {3883} (\bibinfo {year} {2011})}\BibitemShut
  {NoStop}%
\bibitem [{Note2()}]{Note2}%
  \BibitemOpen
  \bibinfo {note} {The only effect of MoS$_2$ on the low-energy states we see
  in a close-up is a small band gap of less than 1 meV respectively 2 meV with
  Re impurities. This is supposed to be analogous to the case of graphene on
  boron nitride, where small finite band gaps can occur due to local mass terms
  in the moir\'{e} cell \cite {sachs2011hBN}. However, this gap is too small to
  limit electron mobility significantly.}\BibitemShut {Stop}%
\bibitem [{\citenamefont {Ristein}, \citenamefont {Mammadov},\ and\
  \citenamefont {Seyller}(2012)}]{SiCdoping}%
  \BibitemOpen
  \bibfield  {author} {\bibinfo {author} {\bibfnamefont {J.}~\bibnamefont
  {Ristein}}, \bibinfo {author} {\bibfnamefont {S.}~\bibnamefont {Mammadov}}, \
  and\ \bibinfo {author} {\bibfnamefont {T.}~\bibnamefont {Seyller}},\ } {\bibfield  {journal} {\bibinfo
  {journal} {Phys. Rev. Lett.}\ }\textbf {\bibinfo {volume} {108}},\ \bibinfo
  {pages} {246104} (\bibinfo {year} {2012})}\BibitemShut {NoStop}%
\bibitem [{\citenamefont {Kim}\ \emph {et~al.}(2008)\citenamefont {Kim},
  \citenamefont {Ihm}, \citenamefont {Choi},\ and\ \citenamefont
  {Son}}]{kim2008origin}%
  \BibitemOpen
  \bibfield  {author} {\bibinfo {author} {\bibfnamefont {S.}~\bibnamefont
  {Kim}}, \bibinfo {author} {\bibfnamefont {J.}~\bibnamefont {Ihm}}, \bibinfo
  {author} {\bibfnamefont {H.~J.}~\bibnamefont {Choi}}, \ and\ \bibinfo {author}
  {\bibfnamefont {Y.-W.}\ \bibnamefont {Son}},\ }\href@noop {} {\bibfield
  {journal} {\bibinfo  {journal} {Phys. Rev. Lett.}\ }\textbf {\bibinfo
  {volume} {100}},\ \bibinfo {pages} {176802} (\bibinfo {year}
  {2008})}\BibitemShut {NoStop}%
\bibitem [{\citenamefont {Stellman}(1998)}]{stellman1998encyclopaedia}%
  \BibitemOpen
  \bibfield  {author} {\bibinfo {author} {\bibfnamefont {J.}~\bibnamefont
  {Stellman}},\ }\href@noop {} {\emph {\bibinfo {title} {Encyclopaedia of
  occupational health and safety}}}\ (\bibinfo  {publisher} {Intl Labour
  Organisation},\ \bibinfo {year} {1998})\BibitemShut {NoStop}%
\bibitem [{\citenamefont {Greenwood}\ and\ \citenamefont
  {Earnshaw}(1997)}]{Greenwooed1997chemistry}%
  \BibitemOpen
  \bibfield  {author} {\bibinfo {author} {\bibfnamefont {N.}~\bibnamefont
  {Greenwood}}\ and\ \bibinfo {author} {\bibfnamefont {A.}~\bibnamefont
  {Earnshaw}},\ }\href@noop {} {\emph {\bibinfo {title} {Chemistry of the
  Elements (2nd ed.)}}}\ (\bibinfo  {publisher} {Butterworth-Heinemann},\
  \bibinfo {year} {1997})\BibitemShut {NoStop}%
\bibitem [{\citenamefont {Bollinger}\ \emph {et~al.}(2001)\citenamefont
  {Bollinger}, \citenamefont {Lauritsen}, \citenamefont {Jacobsen},
  \citenamefont {N\o{}rskov}, \citenamefont {Helveg},\ and\ \citenamefont
  {Besenbacher}}]{edges1}%
  \BibitemOpen
  \bibfield  {author} {\bibinfo {author} {\bibfnamefont {M.~V.}\ \bibnamefont
  {Bollinger}}, \bibinfo {author} {\bibfnamefont {J.~V.}\ \bibnamefont
  {Lauritsen}}, \bibinfo {author} {\bibfnamefont {K.~W.}\ \bibnamefont
  {Jacobsen}}, \bibinfo {author} {\bibfnamefont {J.~K.}\ \bibnamefont
  {N\o{}rskov}}, \bibinfo {author} {\bibfnamefont {S.}~\bibnamefont {Helveg}},
  \ and\ \bibinfo {author} {\bibfnamefont {F.}~\bibnamefont {Besenbacher}},\
  } {\bibfield  {journal}
  {\bibinfo  {journal} {Phys. Rev. Lett.}\ }\textbf {\bibinfo {volume} {87}},\
  \bibinfo {pages} {196803} (\bibinfo {year} {2001})}\BibitemShut {NoStop}%
\bibitem [{\citenamefont {Bollinger}, \citenamefont {Jacobsen},\ and\
  \citenamefont {N\o{}rskov}(2003)}]{edges2}%
  \BibitemOpen
  \bibfield  {author} {\bibinfo {author} {\bibfnamefont {M.~V.}\ \bibnamefont
  {Bollinger}}, \bibinfo {author} {\bibfnamefont {K.~W.}\ \bibnamefont
  {Jacobsen}}, \ and\ \bibinfo {author} {\bibfnamefont {J.~K.}\ \bibnamefont
  {N\o{}rskov}},\ } {\bibfield
  {journal} {\bibinfo  {journal} {Phys. Rev. B}\ }\textbf {\bibinfo {volume}
  {67}},\ \bibinfo {pages} {085410} (\bibinfo {year} {2003})}\BibitemShut
  {NoStop}%
\bibitem [{\citenamefont {Khomyakov}\ \emph {et~al.}(2010)\citenamefont
  {Khomyakov}, \citenamefont {Starikov}, \citenamefont {Brocks},\ and\
  \citenamefont {Kelly}}]{Khomyakov_gr_metals}%
  \BibitemOpen
  \bibfield  {author} {\bibinfo {author} {\bibfnamefont {P.~A.}\ \bibnamefont
  {Khomyakov}}, \bibinfo {author} {\bibfnamefont {A.~A.}\ \bibnamefont
  {Starikov}}, \bibinfo {author} {\bibfnamefont {G.}~\bibnamefont {Brocks}}, \
  and\ \bibinfo {author} {\bibfnamefont {P.~J.}\ \bibnamefont {Kelly}},\ } {\bibfield  {journal} {\bibinfo
  {journal} {Phys. Rev. B}\ }\textbf {\bibinfo {volume} {82}},\ \bibinfo
  {pages} {115437} (\bibinfo {year} {2010})}\BibitemShut {NoStop}%
\bibitem [{\citenamefont {Park}, \citenamefont {Ahn},\ and\ \citenamefont
  {Ruiz-Vargas}(2009)}]{park2009photocurrent}%
  \BibitemOpen
  \bibfield  {author} {\bibinfo {author} {\bibfnamefont {J.}~\bibnamefont
  {Park}}, \bibinfo {author} {\bibfnamefont {Y.~H.}\ \bibnamefont {Ahn}}, \
  and\ \bibinfo {author} {\bibfnamefont {C.}~\bibnamefont {Ruiz-Vargas}},\
  }\href@noop {} {\bibfield  {journal} {\bibinfo  {journal} {Nano Lett.}\
  }\textbf {\bibinfo {volume} {9}},\ \bibinfo {pages} {1742} (\bibinfo {year}
  {2009})}\BibitemShut {NoStop}%
\bibitem [{\citenamefont {Xia}\ \emph {et~al.}(2009)\citenamefont {Xia},
  \citenamefont {Mueller}, \citenamefont {Golizadeh-Mojarad}, \citenamefont
  {Freitag}, \citenamefont {Lin}, \citenamefont {Tsang}, \citenamefont
  {Perebeinos},\ and\ \citenamefont {Avouris}}]{xia2009photocurrent}%
  \BibitemOpen
  \bibfield  {author} {\bibinfo {author} {\bibfnamefont {F.}~\bibnamefont
  {Xia}}, \bibinfo {author} {\bibfnamefont {T.}~\bibnamefont {Mueller}},
  \bibinfo {author} {\bibfnamefont {R.}~\bibnamefont {Golizadeh-Mojarad}},
  \bibinfo {author} {\bibfnamefont {M.}~\bibnamefont {Freitag}}, \bibinfo
  {author} {\bibfnamefont {Y.}~\bibnamefont {Lin}}, \bibinfo {author}
  {\bibfnamefont {J.}~\bibnamefont {Tsang}}, \bibinfo {author} {\bibfnamefont
  {V.}~\bibnamefont {Perebeinos}}, \ and\ \bibinfo {author} {\bibfnamefont
  {P.}~\bibnamefont {Avouris}},\ }\href@noop {} {\bibfield  {journal} {\bibinfo
   {journal} {Nano Lett.}\ }\textbf {\bibinfo {volume} {9}},\ \bibinfo {pages}
  {1039} (\bibinfo {year} {2009})}\BibitemShut {NoStop}%
\bibitem [{\citenamefont {Echtermeyer}\ \emph {et~al.}(2011)\citenamefont
  {Echtermeyer}, \citenamefont {Britnell}, \citenamefont {Jasnos},
  \citenamefont {Lombardo}, \citenamefont {Gorbachev}, \citenamefont
  {Grigorenko}, \citenamefont {Geim}, \citenamefont {Ferrari},\ and\
  \citenamefont {Novoselov}}]{echtermeyer2011strong}%
  \BibitemOpen
  \bibfield  {author} {\bibinfo {author} {\bibfnamefont {T.~J.}\ \bibnamefont
  {Echtermeyer}}, \bibinfo {author} {\bibfnamefont {L.}~\bibnamefont
  {Britnell}}, \bibinfo {author} {\bibfnamefont {P.~K.}\ \bibnamefont
  {Jasnos}}, \bibinfo {author} {\bibfnamefont {A.}~\bibnamefont {Lombardo}},
  \bibinfo {author} {\bibfnamefont {R.~V.}\ \bibnamefont {Gorbachev}}, \bibinfo
  {author} {\bibfnamefont {A.~N.}\ \bibnamefont {Grigorenko}}, \bibinfo
  {author} {\bibfnamefont {A.~K.}\ \bibnamefont {Geim}}, \bibinfo {author}
  {\bibfnamefont {A.~C.}\ \bibnamefont {Ferrari}}, \ and\ \bibinfo {author}
  {\bibfnamefont {K.~S.}\ \bibnamefont {Novoselov}},\ }\href@noop {} {\bibfield
   {journal} {\bibinfo  {journal} {Nat. Commun.}\ }\textbf {\bibinfo {volume}
  {2}},\ \bibinfo {pages} {458} (\bibinfo {year} {2011})}\BibitemShut {NoStop}%
\bibitem [{\citenamefont {Lee}\ \emph {et~al.}(2008)\citenamefont {Lee},
  \citenamefont {Balasubramanian}, \citenamefont {Weitz}, \citenamefont
  {Burghard},\ and\ \citenamefont {Kern}}]{lee2008contact}%
  \BibitemOpen
  \bibfield  {author} {\bibinfo {author} {\bibfnamefont {E.~J.~H.}\
  \bibnamefont {Lee}}, \bibinfo {author} {\bibfnamefont {K.}~\bibnamefont
  {Balasubramanian}}, \bibinfo {author} {\bibfnamefont {R.~T.}\ \bibnamefont
  {Weitz}}, \bibinfo {author} {\bibfnamefont {M.}~\bibnamefont {Burghard}}, \
  and\ \bibinfo {author} {\bibfnamefont {K.}~\bibnamefont {Kern}},\ }\href@noop
  {} {\bibfield  {journal} {\bibinfo  {journal} {Nature Nanotech.}\ }\textbf
  {\bibinfo {volume} {3}},\ \bibinfo {pages} {486} (\bibinfo {year}
  {2008})}\BibitemShut {NoStop}%
\bibitem [{\citenamefont {Zhang}\ \emph {et~al.}(2013)\citenamefont {Zhang},
  \citenamefont {Chuu}, \citenamefont {Huang}, \citenamefont {Chen},
  \citenamefont {Tsai}, \citenamefont {Chang}, \citenamefont {Liang},
  \citenamefont {He}, \citenamefont {Chou},\ and\ \citenamefont
  {Li}}]{zhang2013ultrahigh}%
  \BibitemOpen
  \bibfield  {author} {\bibinfo {author} {\bibfnamefont {W.}~\bibnamefont
  {Zhang}}, \bibinfo {author} {\bibfnamefont {C.-P.}\ \bibnamefont {Chuu}},
  \bibinfo {author} {\bibfnamefont {J.-K.}\ \bibnamefont {Huang}}, \bibinfo
  {author} {\bibfnamefont {C.-H.}\ \bibnamefont {Chen}}, \bibinfo {author}
  {\bibfnamefont {M.-L.}\ \bibnamefont {Tsai}}, \bibinfo {author}
  {\bibfnamefont {Y.-H.}\ \bibnamefont {Chang}}, \bibinfo {author}
  {\bibfnamefont {C.-T.}\ \bibnamefont {Liang}}, \bibinfo {author}
  {\bibfnamefont {J.-H.}\ \bibnamefont {He}}, \bibinfo {author} {\bibfnamefont
  {M.-Y.}\ \bibnamefont {Chou}}, \ and\ \bibinfo {author} {\bibfnamefont
  {L.-J.}\ \bibnamefont {Li}},\ }\href@noop {} {\bibfield  {journal} {\bibinfo
  {journal} {arXiv preprint arXiv:1302.1230}\ } (\bibinfo {year}
  {2013})}\BibitemShut {NoStop}%
\bibitem [{\citenamefont {Sachs}\ \emph {et~al.}(2011)\citenamefont {Sachs},
  \citenamefont {Wehling}, \citenamefont {Katsnelson},\ and\ \citenamefont
  {Lichtenstein}}]{sachs2011hBN}%
  \BibitemOpen
  \bibfield  {author} {\bibinfo {author} {\bibfnamefont {B.}~\bibnamefont
  {Sachs}}, \bibinfo {author} {\bibfnamefont {T.~O.}\ \bibnamefont {Wehling}},
  \bibinfo {author} {\bibfnamefont {M.~I.}\ \bibnamefont {Katsnelson}}, \ and\
  \bibinfo {author} {\bibfnamefont {A.~I.}\ \bibnamefont {Lichtenstein}},\
  }\href@noop {} {\bibfield  {journal} {\bibinfo  {journal} {Phys. Rev. B}\
  }\textbf {\bibinfo {volume} {84}},\ \bibinfo {pages} {195414} (\bibinfo
  {year} {2011})}\BibitemShut {NoStop}%
\end{thebibliography}
\end{document}

\title{Supplementary Material: Doping Mechanisms in Graphene-MoS$_2$ Hybrids}

\author{B. Sachs}

\affiliation{I. Institut f{\"u}r Theoretische Physik, Universit{\"a}t Hamburg, Jungiusstra{\ss}e 9, D-20355 Hamburg, Germany}

\email{bsachs@physnet.uni-hamburg.de}

\author{L. Britnell}

\affiliation{School of Physics and Astronomy, University of Manchester, M13 9PL Manchester, United Kingdom}

\author{T. O. Wehling}

\affiliation{Institut f{\"u}r Theoretische Physik, Universit{\"a}t Bremen, Otto-Hahn-Allee 1, D-28359 Bremen, Germany}

\affiliation{Bremen Center for Computational Materials Science, Universit{\"a}t Bremen, Am Fallturm 1a, D-28359 Bremen, Germany}

\author{A. Eckmann}

\affiliation{School of Physics and Astronomy, University of Manchester, M13 9PL Manchester, United Kingdom}

\author{R. Jalil}

\affiliation{Manchester Centre for Mesoscience and Nanotechnology, University of Manchester, Manchester M13 9PL, United Kingdom}

\author{B. D. Belle}

\affiliation{Manchester Centre for Mesoscience and Nanotechnology, University of Manchester, Manchester M13 9PL, United Kingdom}

\author{A. I. Lichtenstein}

\affiliation{I. Institut f{\"u}r Theoretische Physik, Universit{\"a}t Hamburg, Jungiusstra{\ss}e 9, D-20355 Hamburg, Germany}

\author{M. I. Katsnelson}

\affiliation{Radboud University of Nijmegen, Institute for Molecules and Materials, Heijendaalseweg 135, 6525 AJ Nijmegen, The Netherlands}

\author{K. S. Novoselov}

\affiliation{School of Physics and Astronomy, University of Manchester, M13 9PL, Manchester, United Kingdom}

\maketitle

\section{Computational method}

Density functional theory (DFT) calculations were performed using the Vienna ab initio simulations package (VASP) \cite{kresse_vasp} with projector augmented (PAW) plane waves \cite{PAW1,PAW2}. The local density approximation (LDA) was employed to the exchange-correlation potential, which is better suited to describe van der Waals forces in weakly bound layered systems than generalized gradient (GGA) functionals \cite{LDAGGA_Schwarz,sachs2011hBN,hasegawa2004semiempirical}. For simulations of impurities in the heterostructure, as shown in Fig. 1 of our paper, a k-mesh of 12x12x1 points and a plane-wave cut-off of 500 eV was employed.

For simluations of edge effects, the unit cell in Fig. 1a of the paper was repeated 8 times in one direction, whereby only half of the supercell was covered by MoS$_2$. Here, a 3x1x1 k-mesh was selected together with a plane-wave cut-off of 400 eV. Rhenium impurities were included in the same concentration as discussed in Fig. 1b of the paper with a minimum distance of Re impurities to the MoS$_2$ edge of about 10 \AA.

\begin{figure}[htb]
\includegraphics[width=0.8\linewidth]{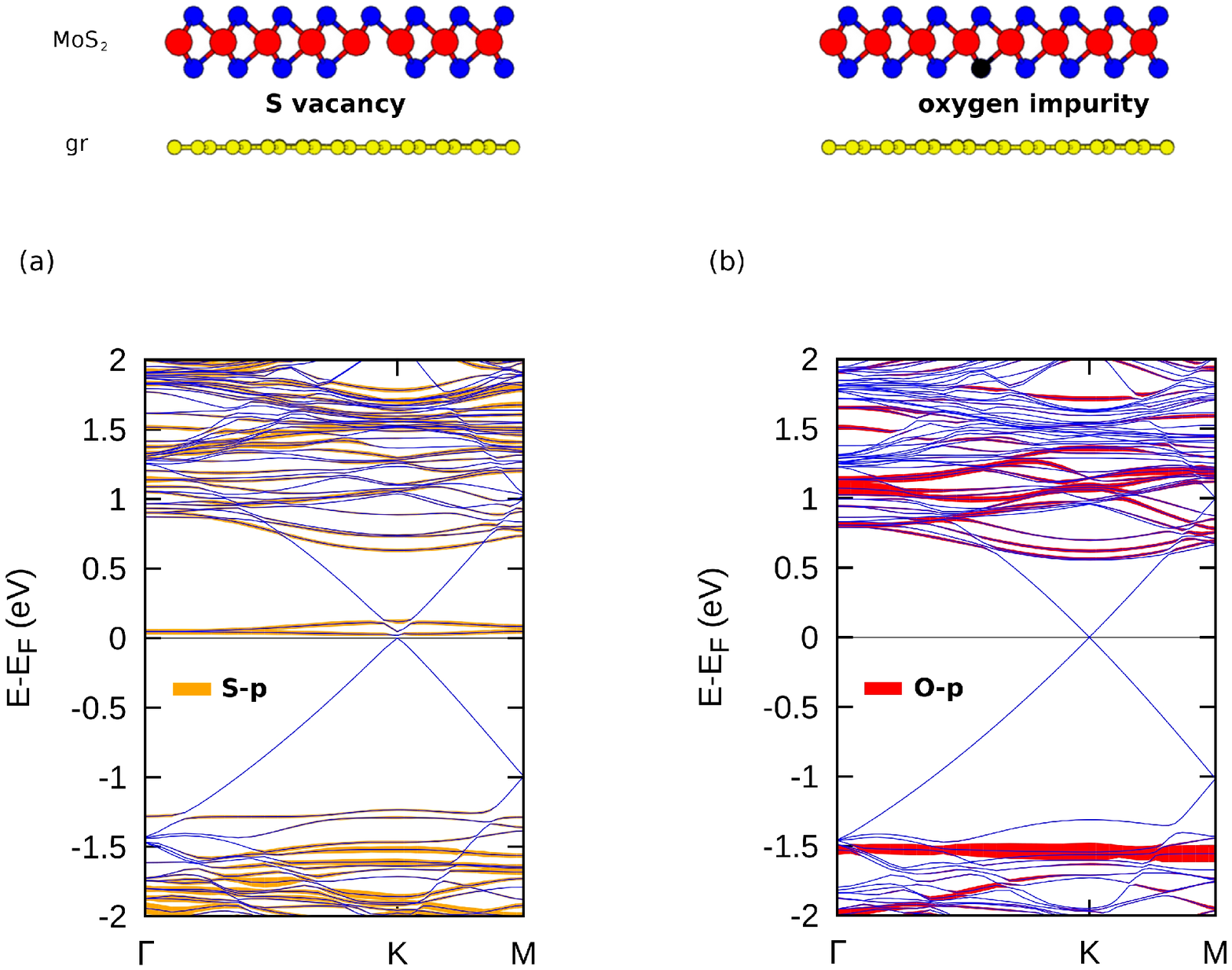}
\caption{(a) (top) Sketch of the graphene-MoS$_2$ system from side view to visualize the simulated sulfur defect position in the system (yellow atoms indicate the graphene carbon atoms, blue atoms sulfur and red atoms molybdenum). (bottom) Band diagram (blue bands). Orange thick bands show $p$ state contributions of S atoms next to the sulfur vacancy. (b) The same for oxygen impurities. Red thick bands show $p$ state contributions of the oxygen impurity.}
\label{fig:bands_def}
\end{figure}

\section{Sample preparation}

The measurements were performed on samples comprised of a graphene - MoS$_2$ heterostructure with Au electrodes contacting the graphene. The samples were prepared by first mechanically cleaving a graphene flake onto a Si/SiO$_2$ (300 nm oxide layer) substrate \cite{graphene_first}. A separately prepared MoS$_2$ flake was then transferred to the surface of the graphene flake. For this, a dry transfer technique \cite{dean2010boron} was used whereby the MoS$_2$ was prepared on a polymer substrate, inverted and placed in the chosen position to ensure a clean interface between the two materials. Au contacts were then fabricated by standard photolithographic processing. 

For measurements of the gate voltage response (Fig.  4d of the paper), a sample with several layers of MoS$_2$ without additional graphene layers on top was used, which explains the small photocurrent compared to the photocurrent maps. We also simulated the effect of MoS$_2$ multilayers on the doping and we found it to be not important (see below).

\section{Sulfur vacancies and oxygen impurities in M\lowercase{o}S$_2$} 

In the paper, we theoretically investigated the scenario of Re impurities in MoS$_2$ and showed that these dope graphene with electrons. However, other impurity types that can possibly occur have to be excluded as a source of the doping. A possible source for a charge transfer between MoS$_2$ and graphene could be sulfur vacancies in MoS$_2$. We simulated sulfur vacancies in the spirit of Fig. 1 in the paper using the same geometry, but by removing one out of 32 sulfur atoms with further relaxation of the structure. The resulting band diagram is shown in Fig. \ref{fig:bands_def}a. First, we can clearly see that graphene's Dirac cone is perfectly aligned with the Fermi level, so graphene is undoped and there is no charge transfer between graphene and defective MoS$_2$ with S vacancies, even in this high concentration. The main difference to the pristine MoS$_2$/graphene interface can be found close to the Fermi level: here, it can be seen that hole states emerge from $p$ states of S atoms sitting next to the vacancy (orange thick bands).

\begin{figure}[htb]
\includegraphics[width=0.4\linewidth]{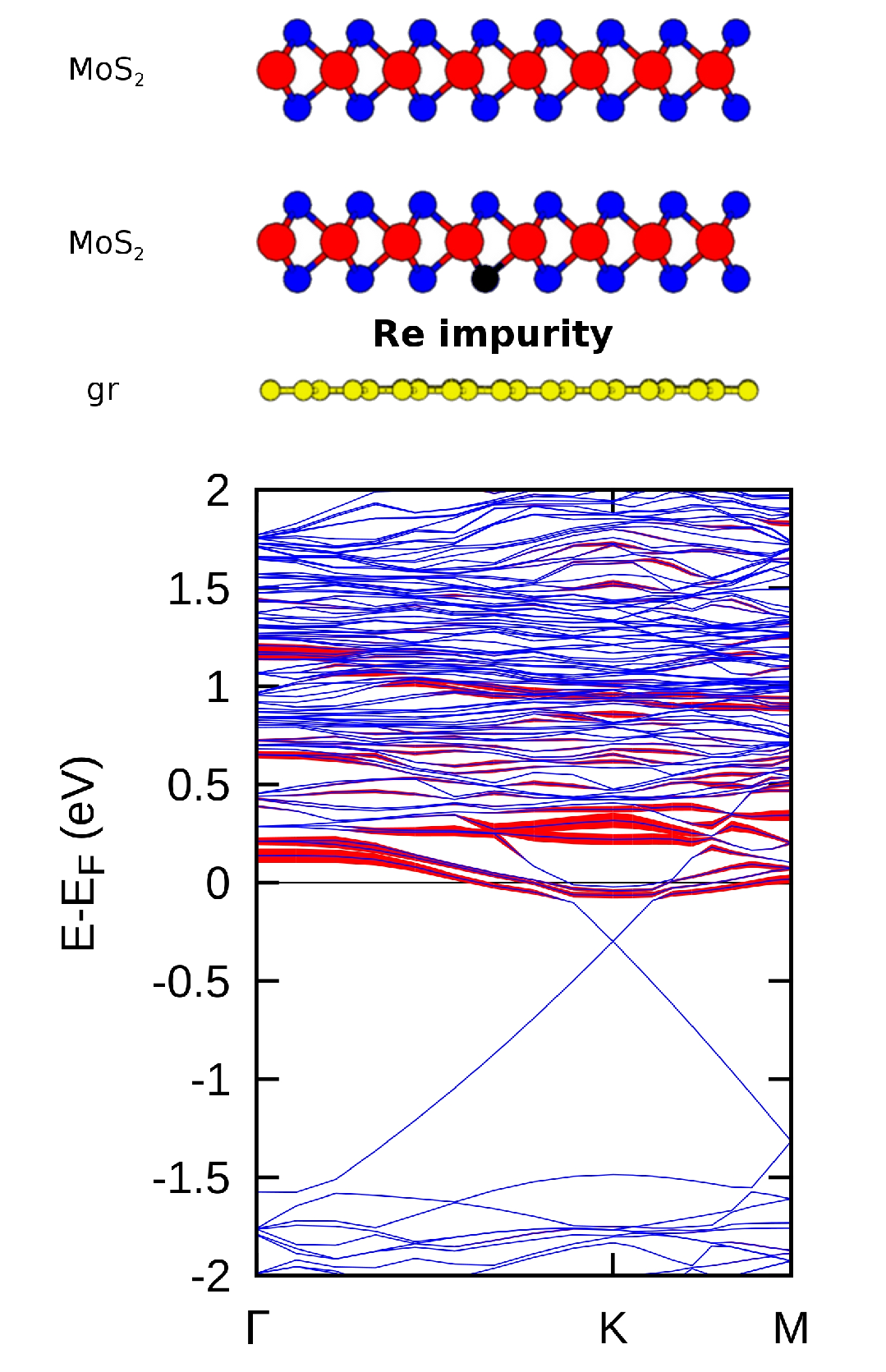}
\caption{(a) (top) Sketch of the graphene-MoS$_2$ system from side view with two MoS$_2$ layers used (yellow atoms indicate the graphene carbon atoms, blue atoms sulfur and red atoms molybdenum). (bottom) Band diagram (blue bands). The red thick bands visualize contributions of the Re $d$ orbitals to the band structure.}
\label{fig:twoMoS2}
\end{figure}

Since the XPS measurements exhibited a significant amount of oxygen in the samples, we also simulated MoO$_x$S$_{2-x}$ in the heterostructure to exclude doping effects here. Again, the same unit cell as in Fig. 1a of the manuscript was used, but now with one S atom replaced by an oygen atom, so $x=0.0625$ (Fig. \ref{fig:bands_def}b). Here, the band diagram looks very similar to the pristine system shown in Fig. 1a in the main text. Since sulfur and oxygen are members of the same group of the periodic table, the covalent bonds of S and O atoms with neighboring Mo atoms are similar. Therefore, the oxygen $p$ states (red colored thick bands) in the contaminated heterostructure do not affect the electronic structure significantly.

\section{M\lowercase{o}S$_2$-thickness and temperature dependence of the doping} 

For photocurrent measurement, MoS$_2$ flakes were used in the heterostructures with a thickness of several layers, while simulations were performed considering a single-layer of MoS$_2$ only. To be sure that the doping of graphene is independent of the MoS$_2$ flake thickness, we repeated the same calculations as shown in Fig. 1b of the paper, but with two layers of MoS$_2$ on top. The resulting band diagram is shown in Fig. \ref{fig:twoMoS2}. We find only a very small change of the Fermi level shift of about 15 meV downwards and the bands look very similar compared to the single-layer graphene/MoS$_2$ heterostructure. Therefore, we conclude that the charge transfer in the interface region is not sensitive to the thickness of the MoS$_2$ flake and the modelled system is well-suited to simulate the doping in a heterostructure with multi-layer MoS$_2$.

We also tested the temperature influence on the Fermi level position in our calculations by extracting the carrier density from density of states calculations, which we weighted by Fermi-Dirac distribution. For measurements performed at room temperature, temperature-induced Fermi level shifts are on the order of 10meV and therefore negligible.

\section{Charge reordering in graphene under different M\lowercase{o}S$_2$ edge types}

\begin{figure}[htb]
\includegraphics[width=0.6\linewidth]{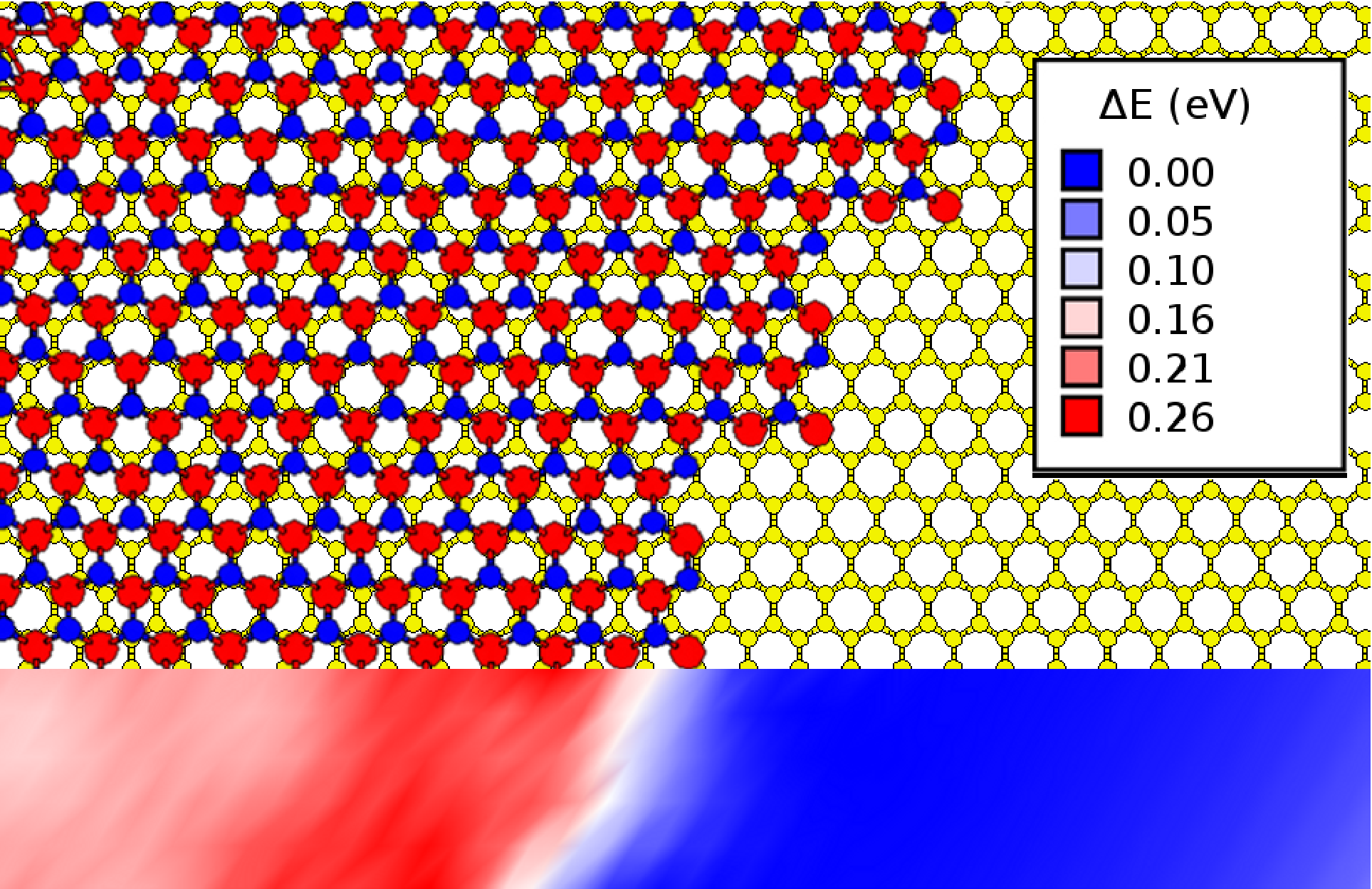}
\caption{Top view on the MoS$_2$-graphene heterostructure (yellow atoms indicate the graphene carbon atoms, blue atoms sulfur and red atoms molybdenum). The contour plot shows the on-site energy differences of graphene $p_{\rm z}$ orbitals.}
\label{fig:edgearm}
\end{figure}

In Fig. 3 of our paper, we discuss the reordering of charge in graphene below a MoS$_2$ edge for the case of a (\=1010) edge. We also investigated (10\=10) edge types. We find that the shape of the potential step in graphene below MoS$_2$ varies with these realistic edge types \cite{edges1,edges2}. However, there are important similarities in the potential step for all edge types. Neglecting additional impurities, the absolute value of the potential step varies, but is always on the order of some hundred meV. Most importantly, the resulting electric field is always directed such that electrons move to the uncovered regions while holes migrate to the covered part of the edge region. We also tested a complex edge structure with lateral cuts in zigzag and armchair direction. This edge type is artificial because we did not saturate edge atoms and did not relax the structure. The structure is shown from top view in Fig. \ref{fig:edgearm}. Although armchair and zigzag states in MoS$_2$ nanoribbons were shown to be electronically very different \cite{erdogan2012transport}, we see from the color map of the graphene $p_{\rm z}$ on-site energies ($E= \int_{-\infty}^{E_{\rm F}}{\epsilon\rho\left(\epsilon \right)d\epsilon}/\int_{-\infty}^{E_{\rm F}}{d\epsilon}$, with $\rho\left(\epsilon \right)$ the carbon $p_{\rm z}$-density of states) again a potential step comparable to the steps under other edge types. 

Hence, we always find a potential step in graphene below MoS$_2$ edges, and the potential landscape becomes even more manifold in the presence of impurities which we show in Fig. 3b of the paper. Although the exact local shape of the potential landscape in the experiment is not known, we can conclude from our simulations that the photocurrent is generated at these edges which is supported by symmetries in the photocurrent maps.